\documentclass[12pt,preprint]{aastex}

\newcommand{\kms}{\mbox{km s$^{-1}$}}

\newcommand{\ha}{\mbox{H$\alpha$}}
\newcommand{\hb}{\mbox{H$\beta$}}

\newcommand{\lam}{\mbox{$\lambda$}}

\newcommand{\msun}{\mbox{M$_{\odot}$}}

\shorttitle{Small scale dynamics of a $z\sim3$ LBG}

\begin{document}

\title{Lyman Break Galaxies Under a Microscope: The Small Scale
Dynamics and Mass of an Arc in the Cluster 1E0657-56\altaffilmark{1}}

\author{N.P.H. Nesvadba\altaffilmark{2}, M.D. Lehnert\altaffilmark{2},
F. Eisenhauer\altaffilmark{2}, R. Genzel\altaffilmark{2},
S. Seitz\altaffilmark{2}, R.I. Davies\altaffilmark{2},
R.P. Saglia\altaffilmark{2}, D. Lutz\altaffilmark{2}, 
L. Tacconi\altaffilmark{2}, R. Bender\altaffilmark{2},
R. Abuter\altaffilmark{2}}  

\altaffiltext{1}{Based on observations collected at the European Southern
Observatory, Very Large Telescope Array, Cerro Paranal;
program numbers, 70.B-0545(A) and 70.A-0229(A)}

\altaffiltext{2}{Max-Planck-Institut f\"ur extraterrestrische Physik,
Giessenbachstra\ss e, 85748 Garching bei M\"unchen, Germany}

\begin{abstract} 
Using the near-infrared integral-field spectrograph SPIFFI on the VLT,
we have studied the spatially-resolved dynamics in the $z=3.2$ strongly
lensed galaxy 1E0657-56 ``arc$+$core'' by observing the rest-frame
optical emission lines [OIII]$\lambda$5007 and H$\beta$. The lensing
configuration suggests that the high surface brightness ``core'' is the
${\cal M}\sim 20$ magnified central $\sim 1$ kpc of the galaxy, whereas
the fainter ``arc'' is the more strongly magnified peripheral region of
the same galaxy at about a half-light radius, which otherwise appears
to be a typical $z\sim 3$ Lyman break galaxy.

The overall shape of the position-velocity diagram resembles the
``rotation curves'' of the inner few kpcs of nearby $\sim{\cal L}^*$
spiral galaxies. For ${\cal M}=20$, our data have a spatial resolution of
$\sim$200 pc in the source plane. The projected velocities $v_{rot,proj}$
rise rapidly to $\sim 75$ \kms\ within radii $\sim 0.5$ kpc from the
center, and asymptotically reach a velocity of $\sim 190$ \kms\ within
the arc, at a projected radius of a few kpc radius. The rotation curve
implies a dynamical mass of $\log{M_{dyn}/M_{\odot}}\sim 9.3$ within the
central kpc, and suggests that in this system the equivalent of the mass
of a present-day $\sim{\cal L}^*$ bulge at the same radius was already in
place by $z\gtrsim 3$. Approximating the circular velocity of the halo
by the measured asymptotic velocity of the rotation curve, we estimate
a dark matter halo mass of $\log{M_{halo}/M_{\odot}}\sim 11.7\pm 0.3$,
in good agreement with large-scale clustering studies of Lyman break
galaxies. The baryonic collapse fraction is low compared to $z\sim 2$
actively star-forming ``BX'' and low-redshift galaxies, perhaps implying
comparatively less gas infall to small radii or efficient feedback. Even
more speculatively, the high central mass density might indicate highly
dissipative gas collapse in very early stages of galaxy evolution, in
approximate agreement with what is expected for ``inside-out'' galaxy
formation models.

\end{abstract}


\section{Introduction}

Dynamical mass estimates of high redshift galaxies are now starting
to play a significant role in our developing understanding of galaxy
assembly in the early universe, a trend that will likely become even
more important in the near future. Directly measuring the dynamical
masses from the spatially-resolved spectra of high-redshift galaxies
is observationally very challenging, but dynamical masses are less
prone to degeneracies and evolutionary bias than mass estimates
based solely on photometry. Fascinatingly, they may also allow us to
directly probe the baryonic and dark matter content and concentration of
galaxies in the early universe, and to measure their angular momenta 
\citep{NMFS06}. Ultimately, they will enable us, for example, to
directly compare the growth of galaxy mass and angular momentum with
model predictions as a function of redshift. Accurately measuring the
kinematics of high-redshift galaxies is therefore a major step forward.

To realize these goals, we must show convincingly that the kinematics we
measure in a high redshift galaxy have a simple proportionality to the
mass distribution and rule out that they are dominated by the orbit or
angular momentum loss of mergers or by hydrodynamical processes like,
e.g., starburst-driven ``superwinds'' \citep{lehnert96a}.

The Lyman-break technique has led to the largest sample of
spectroscopically confirmed galaxies from z$\sim$2.7 to 6.4. Despite
our rapidly growing understanding of their ensemble properties, such
as their luminosity function, clustering, and star-formation history
\citep[e.g.,][]{steidel96, adelberger98}, our knowledge of their detailed
intrinsic properties remains rather rudimentary.  LBGs at z$\sim$3
have typical radii of $r_{e} \sim 0.3\arcsec$ \citep{giavalisco96}
so that the spatially resolved kinematics are often difficult to
obtain. Thus, dynamical mass estimates for individual LBGs at z$\sim$3
\citep[e.g.,][]{pettini01} are based mostly on line widths and only
in a handful of cases have velocity gradients been observed. Spatially
resolved LBGs are large compared to the overall population, and might
be biased towards the strongest line emitting galaxies (e.g., vigorous
starbursts) or early-stage mergers and perhaps are not representative
for the overall population.

The only way to properly address these issues is to resolve
the dynamics of an LBG on fine scale.  Strongly gravitationally lensed
LBGs are a promising way to probe small physical scales even with
seeing-limited data. The ideal target would be a strongly-lensed,
highly inclined 
LBG, where the kinematic major axis is roughly along a caustic.  Such a
configuration would allow several patches of the same galaxy, but at
different radii, to be highly magnified to include the intrinsically
low surface-brightness periphery. Probing non-lensed LBGs in this way
is impossible given their generally small radii, faint magnitudes,
and low surface brightnesses.

Unfortunately, strongly lensed LBGs with a favorable lensing geometry
are exceedingly rare.  Two cases at z$>$1 have been studied in detail so
far, MS1512-cB58 at z=2.8 \citep[e.g.,][]{teplitz00, pettini02, baker04}
and AC114-S2 at z=1.9 \citep{L-B03}. However, the underlying dynamical
mechanism is not conclusively revealed in either case. MS1512-cB58 is
magnified by a factor $\sim 30$, but it is compact and has no apparent
velocity gradient \citep{teplitz00} -- most likely because of an
unfavourable lensing geometry. From the mm CO emission line width,
\citet{baker04} measure $M_{dyn} \sim 10^{10}$ \msun, not corrected
for inclination. AC114-S2 has a velocity gradient \citep{L-B03},
but is a merger with complex morphology, and its nature is not well
constrained. That spatially-resolved spectroscopy of giant arcs can
provide valuable constraints on the internal dynamics of the lensed
galaxies has recently been shown by \citet{swinbank06} for a sample of
6 giant arcs at lower redshift, z$\sim$1. \citeauthor{swinbank06}
found regular kinematics in 4 of the 6 galaxies, consistent
with quiescently rotating disks, while in 2 galaxies, they observed
complex line profiles of varying widths and irregular velocity structure
suggestive of either mergers or outflows.

The strongly lensed z=3.24\footnote{We adopt a flat concordance cosmology
with $\Omega_{\Lambda}$=0.7 and H$_0$=70 km s$^{-1}$ Mpc$^{-1}$, in
which D$_L$=27.9 Gpc and D$_A$=1.5 Gpc at $z=3.24$. The size scale is 7.5
kpc/\arcsec. The age of the universe at this redshift and cosmological
model is 1.9~Gyrs.} ``arc$+$core'' galaxy behind the z=0.3 X-ray cluster
1E0657-56 \citep{tucker98} appears to be different from these well-studied
high-redshift gravitational arcs. Its lensing configuration suggests the
simultaneous magnification of a high surface brightness region at the
south-eastern tip of the source that may be associated with the ``core''
of the galaxy as well as a more highly magnified, lower surface brightness
region outside the core \citep[``arc'';][]{mehlert01}.  The total
extent of the arc is $\sim$14\arcsec, and it has a complex substructure:
\citet{mehlert01} identify 3 faint knots of similar surface brightness
within the arc, each separated by a few arcseconds. They propose that
the central highest surface brightness region of the lensed galaxy, lying
near, but outside the cusp-caustic, is seen as the bright core, whereas
a fainter outer region on one side of the same galaxy, which touches
the cusp-caustic and is split into three merging images, constitutes the
full extent of the arc.  Thus the asymmetric magnified image comprising
the near-nuclear region and peripheral patches originating on one side
of the galaxy.  The high magnification (${\cal M}\ga20$) presents an
excellent opportunity to investigate the properties of a z$\sim$3 galaxy
at different radii with high physical and spatial resolution.

The paper is organized as follows: After presenting observations and
data reduction in \S\ref{sec:obsred}, we turn in \S\ref{sec:acs} to the
rest-frame UV and optical properties, highlighting the high-resolution
ACS morphology. We discuss the spatially-resolved rest-frame emission line
kinematics extracted from three--dimensional data cubes obtained with the
integral-field spectrograph SPIFFI in \S\ref{sec:results}. This includes
a detailed discussion of the 146 km s$^{-1}$ velocity gradient in the
core and its continuation in the arc. In \S\ref{sec:spiral}, we present
a detailed comparison with the internal dynamics of low-redshift spiral
galaxies. In \S\ref{sec:evolution}, we estimate the evolution and angular
momentum of the arc$+$core, before investigating the halo mass and the
baryonic ``collapse fraction'' in \S\ref{sec:dmhalo}. We summarize our
results in \S\ref{sec:summary} and draw a likely evolutionary scenario
for LBGs.

\section{Observations and Data Reduction}
\label{sec:obsred}

Given the interesting lensing configuration of the 1E0657-56 arc$+$core
galaxy and the wealth of supplementary data, we observed it with the
near-infrared integral field spectrograph SPIFFI \citep{eisenhauer00},
using UT2 of the VLT (SPIFFI has since become part of the SINFONI
instrument on UT4).
We obtained deep K band spectroscopy of 1E0657-56 
arc core, covering the core and neighboring parts of the arc. Observations
in April 2003 were carried out under variable sky conditions, with a total
integration time of 190 minutes.  One ``off'' frame at a sky position was
taken for each ``on'' frame in an off-on-on-off mode, with a spectral
resolution of R $\sim 2400$ at $2.2\,{\rm \mu m}$ and using the scale
of $0.25\arcsec$ pixel$^{-1}$. Individual exposure times are 600s.

Data reduction was performed extending the package of the standard
IRAF \citep{tody93} tools for reducing longslit spectra. Individual
exposures were dark frame subtracted and flat-fielded using exposures of
an internal calibration lamp.  We identified bad pixels based on dark and
flat-field frames, and replaced them by interpolations of the
surrounding pixels (in all 3 dimensions).  Rectification
and wavelength calibration are done before night sky subtraction, to
account for some spectral flexure between the frames. Curvature in each
individual slit-let of each frame was measured and removed using an arc
lamp, before shifting the spectra to an absolute (vacuum) wavelength
scale with reference to the OH lines in the data.

To account for variations in the night sky emission, we normalize the sky
frame to the average of the object frame separately for each wavelength
before sky subtraction, masking bright foreground objects, and correcting
for residuals of the background subtraction and uncertainties in the
flux calibration by subsequently subtracting the (empty sky) background
separately from each wavelength plane.

The three dimensional data cubes are then reconstructed, assuming that
each slitlet covers exactly 32 pixels.  They are spatially aligned
by cross-correlating the collapsed cubes, and then combined, clipping
deviant pixels. Telluric correction is applied to the combined cube.
Flux scales are obtained from standard star observations. From the light
profile of the standard star, we measure the FWHM spatial resolution to
be $0.6\arcsec\times 0.4\arcsec$\ in right ascension and declination,
respectively.

Data taking and reduction of the rest-frame UV spectroscopy of
the arc and core with {\sc Fors1} on the VLT was described by
\citet{mehlert01}. They also kindly provided their R-band data for the
present work, which has a total exposure time of 3800 s and seeing of
$0.6\arcsec -0.9$\arcsec. \citet{mehlert01} give a full account of how
these data were obtained and reduced.  C. Forman-Jones and collaborators
have recently obtained high-resolution imaging of the cluster 1E0657-56
through the F814W filter using the ACS camera onboard the HST. They kindly
shared with us a protion of their full image containing the region around
the arc$+$core before publication.

\section{High-resolution rest-frame UV Morphology}
\label{sec:acs}

\begin{figure}[htb]
\epsscale{0.5}
\plotone{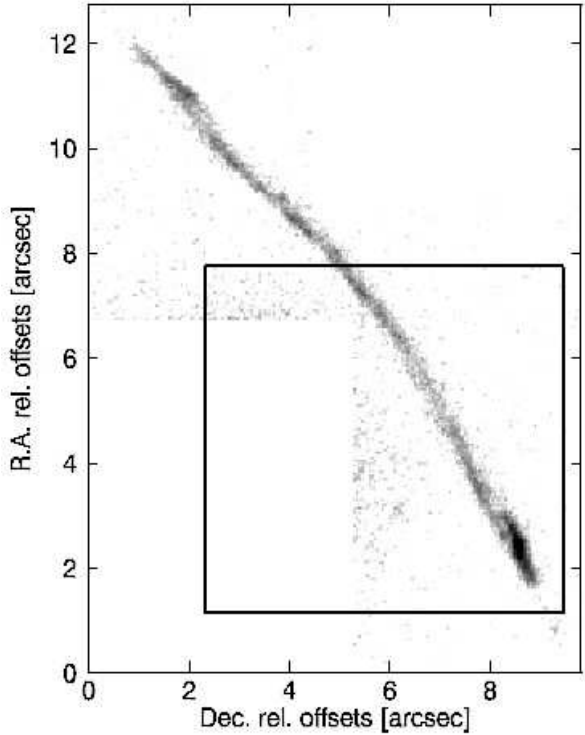}
\caption{{\sc ACS} F814W 
image of the whole arc$+$core.  The core is the relatively high surface
brightness region at the bottom right-hand corner in the image, while the
arc is the lower surface brightness region that extends to the north-east
of the core.  North is at the top, and east is to the left in this image.
The box indicates the {\sc Spiffi} field of view of our data cube. The
ACS data were kindly provided by C. Forman-Jones.}
\label{fig:acsimSpifpos}
\end{figure}

The ACS F814W image ($\sim 2000$\AA\ in the rest-frame) shows the complex
structure of this source in great detail (Fig.~\ref{fig:acsimSpifpos}).
Irregular high surface-brightness patches, perhaps star-forming HII
regions, are seen in the arc, embedded in a more continuous structure
with much lower surface brightness. The core has an overall higher
surface brightness, with a bright, unresolved spot in the center, and
two extensions of lower surface brightness in an S-like shape.

The spatial resolution of the image \citep[obtained from the TinyTim
package;][]{krist97} is 0.09\arcsec\ in both right ascension and
declination. The arc and core are both spatially resolved in the direction
perpendicular to the magnification axis.  Deconvolved profile widths
along the unlensed direction are 0.113\arcsec\ ($\sim$0.9 kpc) for the
brightest part of the core, and $\sim0.3-0.6$\arcsec\ ($2.4 - 4.8$ kpc)
in the arc. Along the magnification axis, the central region of the
core has a diameter of $\sim 0.4$\arcsec, about 4 times larger than the
unlensed 0.113\arcsec\ diameter perpendicular to it.\footnote{Strictly
speaking, this is true only if the gravitational lens has an isothermal
dark matter halo mass profile.  However, the arc$+$core galaxy lies near
the Einstein radius, so the exact profile shape does not have much of an
impact on this estimate.}  The full size of the core along the direction
of magnification is 1.3\arcsec.

\begin{figure}[htb]
\epsscale{0.5}
\plotone{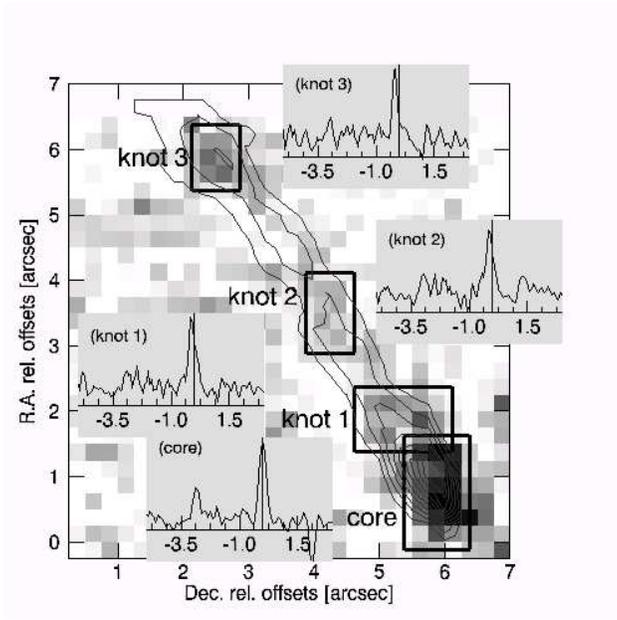}
\caption{Continuum subtracted, combined [OIII]\lam\lam4959,5007 line image
of 1E0657 arc and core {\it (shown as greyscale)}. Contours show the
distribution of the rest-frame UV surface brightness from the R-band
image \citep[kindly provided by][]{mehlert01}. Spectra, extracted from
the rectangular regions as indicated on the image, of areas in the arc
and core are shown in the insets.  The velocities indicated in each
inset are in units of 1000 \kms relative to the core.  The velocity of
the core is indicated by the vertical line in each inset and is based
on the line centroid of the [OIII]\lam5007 emission line within the core.
North is at the top, and east is to the left in this image.}
\label{fig:lineim}
\end{figure}
\section{Spatially-Resolved Spectroscopy and Internal Kinematics} 
\label{sec:results}

Since the lensing direction is roughly along the vertical axis of the
cube, and the SPIFFI data are not spatially resolved in the perpendicular
direction, we simply extract spectra from each individual pixel
row or slitlet which lie along this direction. The core is a bright
[OIII]\lam5007 line emitter, with $S/N=10-20$ in each spectrum, and
uniform dispersions of $\sigma \sim 70$ \kms.  Fitting the centroids of
the [OIII]\lam5007 emission line, we find an overall velocity gradient
of 146 \kms\ over a total physical distance of $0.9 h^{-1}_{70}$ kpc
for ${\cal M} = 20$. Extracting spectra from the 3 knots indicated by
the boxes in Fig.~\ref{fig:lineim} yields offsets relative to the core
of $\Delta v_{knot1} = -104 \pm 8$ \kms, $\Delta v_{knot2}=-152\pm 9$
\kms, and $\Delta v_{knot3} = -239 \pm 7$ \kms. The latter value might
be somewhat influenced by a night sky line residual.\footnote{ We only
state observed velocities, and do not correct for inclination with the
line of sight ($i_{los}$), because of the large uncertainties related to
geometrical distortions of the ${\cal M}=20$ gravitational lens. Moreover,
the lensing implies another correction factor $\sin(i_{lens})$, to account
for the inclination between the kinematic major axis of the arc$+$core
galaxy and the magnification axis. All relative velocities in this
paper are therefore projected velocities, $v_{obs} = \sin(i_{los})\
\sin(i_{lens})\ v_{intrinsic}$.} We will therefore use the average
of $\Delta v_{knot2}$ and $\Delta v_{knot3}$ to estimate the rotation
velocity in the peripheral regions of the arc$+$core galaxy, $v_{rot}\sim
190$ \kms.

\subsection{Evidence for Disk Rotation at $z=3.2$}

We show the velocity curve of the arc$+$core in Fig.~\ref{fig:velcurve}.
Unfortunately, both the cluster potential and influence of nearby galaxies
are not known accurately enough to allow for a robust magnification
estimate.  Thus in Fig.~\ref{fig:velcurve} we show the velocity curve
for a range of assumed magnifications to illustrate this relative
uncertaintly.  Velocities in the arc$+$core decrease monotonically
from South to North in the core (i.e., within the central $\sim$ 1 kpc)
and arc (over a distance of $\sim 6$\arcsec\ or 2.5 kpc). Thanks to the
large magnification and bright line emission in the core, we can trace
the velocity gradient even within the core over 2.3\arcsec, or $\sim 6$
seeing disks, so correlations due to overlapping seeing disks within
the data are negligible. Moreover, this causes the ratio of velocity
gradient $v_{rot}\sim 190$ \kms\ and central velocity dispersion,
$\sigma_{core}\sim 70$ \kms, $v/\sigma \lesssim 3$, to not depend on
seeing.  We overlay the rotation curve for the local galaxy, NGC4419
(Fig.~\ref{fig:velcurve}).  This shows indeed that the velocity curve of
the arc$+$ core is very similar to that of local nearby galaxies like
NGC4419.  NGC4419 was chosen for this purpose because its asymptotic
velocity matches that measured for the arc$+$core particularly well.

\subsection{Alternative Models: The AGN, Wind, and Merger Hypothesis} 

Due to the anisotropic lensing, we cannot use the full two-dimensional
velocity field of the arc$+$core galaxy to distinguish rotation from
alternative models of the origin of the line emission such as an AGN
or starburst driven wind, or a merger of two galaxies. We therefore
base our arguments on the whole of the rest-frame UV \citep{mehlert01}
and optical (this paper) spectral properties.

For a luminous AGN we would expect several characteristic bright
emission lines in the rest-frame UV spectrum, such as NV\lam1240,
SiIV+O IV]\lam1400, CIV\lam1549, C III]+SiIII]\lam1900, etc. Instead
\citet{mehlert01} observe an absorption line spectrum which is typical
of an actively star-forming high-redshift galaxy.
The optical emission lines are narrow, FWHM$\sim 150-160$ \kms\ (see
Table~\ref{tab:lineprops}), with constant (within the uncertainties)
line profiles, line ratios ([OIII]/\hb), and equivalent widths across
both the arc and core. None of these properties provides an indication
of the arc$+$core galaxy hosting a UV/optically bright AGN

If the kinematics of the emission line gas were dominated by superwinds
\citep[e.g.,][]{lehnert96a}, one would not expect the line emission to
have the same morphology as the bright continuum emission.  Line emission
in the arc$+$core galaxy generally follows the continuum morphology,
arguing against the superwind hypothesis. To quantify this, we have
compared the equivalent widths in the arc$+$core with a sample of 12
low redshift, low metallicity galaxies with active star-formation taken
from \citet{storchi95}. Rest-frame \hb\ and [OIII]\lam5007 equivalent
widths in the arc and core are for \hb\ and [OIII]\lam5007, respectively,
W(\hb)$_{arc}$ = 22 \AA, W(\hb)$_{core}$=46 \AA, W([OIII]\lam5007)$_{arc}$
= 89 \AA, W([OIII]\lam5007)$_{core}$ =171 \AA. These values are within
the broad range of equivalent widths measured in the integrated spectra
of the low redshift sample of \citeauthor{storchi95} where the line
emission is not dominated by superwinds.

[OIII]\lam5007/\hb\ line ratios increase by up to 1 dex with increasing
distance from the galactic disk in galaxies exhibiting superwinds due
to the (relative) dominance of shock ionization at large distances from
the disk \citep[e.g.][]{dahlem97, moran99}, even when a strong wind is
projected onto the galaxy continuum \citep[][measure a variation of 0.3
dex in in the low metallicity dwarf galaxy, NGC1569]{devost97}. The arc
and core span a few kpc in the source plane, so that we would expect
a change in line ratios if the emission was arising from the wind. We
find however insignificant differences in [OIII]/\hb\ line ratios,
([OIII]/\hb)$_{arc}=$ 2.53 and ([OIII]/\hb)$_{core}=$ 2.48 in the
arc and core, respectively. This of course does not imply the general
absence of a superwind in this galaxy, it only indicates that the cores
of the optical emission lines are not dominated by an outflow. We find
a similar situation in the z=2.57 strongly star-forming submillimeter
galaxy SMMJ14011+0252 (Nesvadba et al. 2006, in preparation).

To investigate whether the kinematics of the arc$+$core could be due
to a merger, we have constructed a Monte Carlo simulation, evaluating
the likelhood that two unrelated LBGs have similar R-G colors as
that observed by \citet{mehlert01} for the arc and core. To make this
comparison, we used the R$-$G color distribution of \citet{shapley01}
as reference distribution for the colors of LBGs. In 95\% of all
cases, color differences between two random pairs of LBGs are larger
than $\Delta$(R$-$G)$=$0.39 -- the 1$\sigma$ uncertainty of the R$-$G
color difference between the arc and the core. Moreover, the light
distribution appears very smooth and contiguous in the ACS image
with $\lesssim 0.1$\arcsec\ resolution. This means that either
the light profiles of two merging galaxies would have to overall be
very similar, or that their physical separation would have to be less
than $\sim$40 pc. Given the regularity of the surface brightness
distribution in the ACS image (Fig.~\ref{fig:acsimSpifpos}), this
seems highly unlikely. 
The uniform line widths and [OIII]/\hb\ ratios in the arc and core
indicate similar gravitational potentials and overall gas ionization
(excitation and metallicity).  In addition, if this were a merger,
then the absence of obvious irregularities in the velocities and line
widths as a function of projected position would be puzzling given the
high physical resolution due to the strong lensing and the reasonable
number of independent resolution elements across the arc$+$core. 
Moreover, the good agreement of the position velocity diagram with the
rotation curve of NGC4419 requires that the velocity
difference, rotation speeds, positions and relative orientation of
two merging galaxies are very well matched. This fine tuning of
several degrees of freedom makes the merger scenario highly unlikely.

\section{Dynamics and evolutionary stage}
\label{sec:spiral}

\subsection{Dynamical Mass Estimates and Mass Surface Density}

\begin{figure}[htb]
\plotone{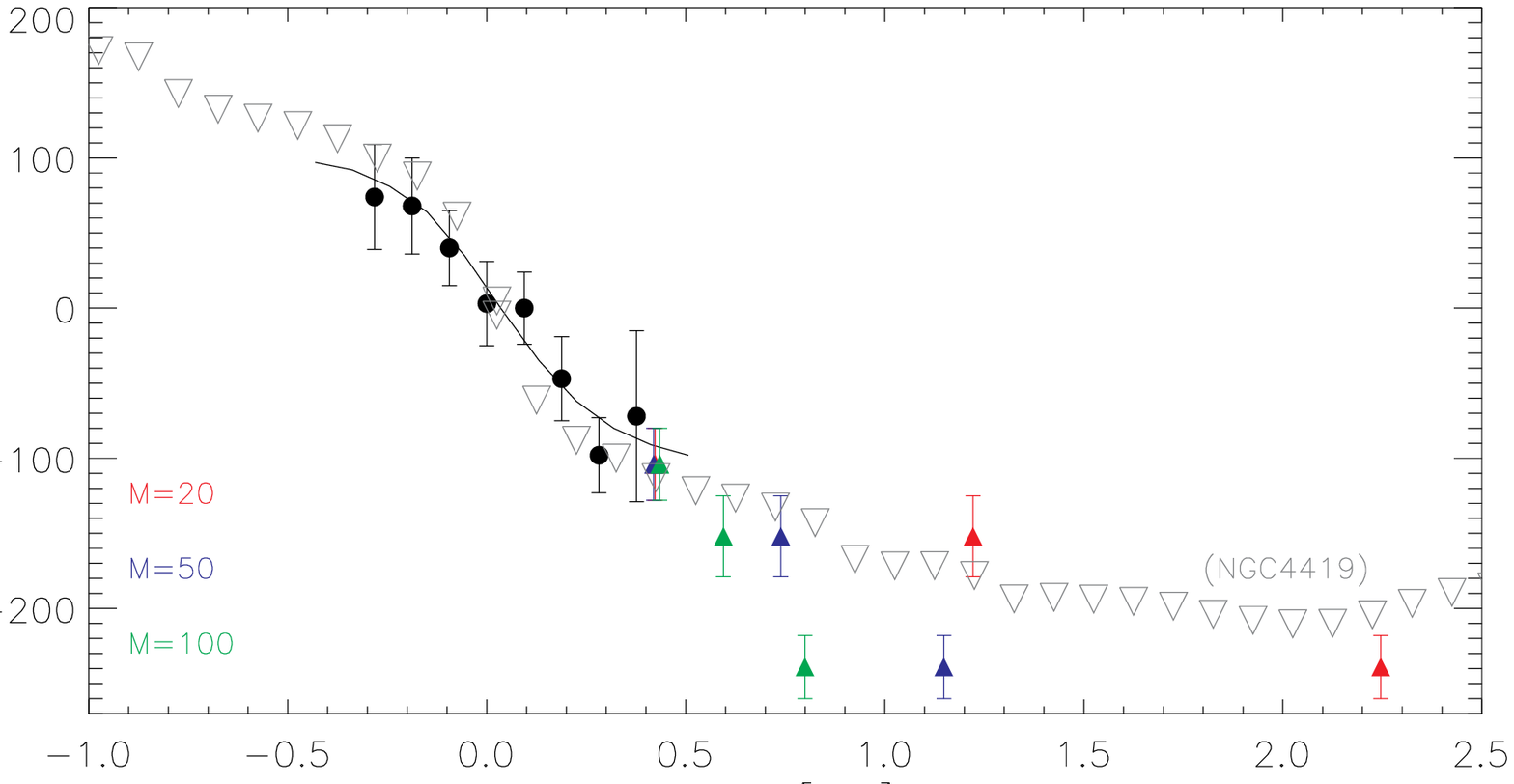}
\caption{[OIII]\lam5007 Position-peak velocity curve of the core and
arc. Data in the core are shown as black dots, triangles indicate
the velocities in the arc for different magnifications (red -- ${\cal
M}=20$, blue -- ${\cal M}=50$, green -- ${\cal M}=100$). Gray upside-down
triangles indicate the digitized rotation curve of NGC4419 taken from
\citet{sofue03}, the black line shows our best-fit elliptical disk model
(for the core only).}
\label{fig:velcurve}
\end{figure}

\citet{sofue03} studied the properties of rotation curves of nearby
disk galaxies and found that low-mass spiral galaxies tend to rotation
speeds that increase out to larger radii (up to several kpc) than do the
more massive disk galaxies. However, the rotation curves of most
galaxies, regardless of mass, 
appear to rise out to a few hundred parsecs and
this rise corresponds roughly to the region consisting of the bulge of
the galaxy.  In Fig.~\ref{fig:velcurve}, we show the high-resolution
CO rotation curve of the nearly edge on ($i\sim 80^{\circ}$) NGC4419,
a dwarf SBa galaxy in the \citet{sofue03} sample with a very similar
position-velocity diagram as the arc$+$core. If the magnification in the
arc does not strongly exceed the ${\cal M} = 20$, then the two curves
agree remarkably well.

Although this excellent agreement is coincidental, it is illustrative
to compare the properties of the arc$+$core galaxy 
and NGC4419. The mass of NGC4419 within $r=0.5$ kpc is $M_{dyn}\sim
1.7 \times 10^{9}$ \msun, whereas at $r=2$ kpc, the enclosed mass is
$M_{dyn}\sim 5\times 10^{9}$ \msun \citep{sofue03}.  As noted earlier,
this does emphasize that interpreting the velocity curve of the arc$+$core
as a rotation curve is justifiable. More specifically, the ACS morphology
and velocity gradient of the arc$+$core galaxy are at least consistent
with the assumption that this is a disk galaxy seen nearly edge-on.

We fit the velocity curve of the core with a simple exponential disk
model, 
accounting for the seeing, magnification by the gravitational lens
(${\cal M} = 20$), and coarse sampling of our data. The simulated data
were extracted in the same way as our observational data. We obtain
a robust fit for a mass of $M_{dyn,0.5 kpc}= 1.3\times 10^9$ \msun\
within a radius $R=0.5$ kpc, assuming an edge-on thin disk parallel to
the lensing axis with no bulge, and similar disk scale-length as the
typical LBG at z$\approx$3. Extrapolating this fit out to $R=2.3$ kpc
\citep[the typical half-light radius of $z\sim 3$ LBGs, ][]{giavalisco96},
we find $M_{dyn,2.3 kpc}\sim 5 \times 10^9$ \msun. Since it is difficult
to place firm constraints on the inclination, we leave it unconstrained,
and only note that, statistically, the mass will be a factor of 2
higher. We also measure a relatively large velocity dispersion in the
core. If the dispersion is indeed due to the gravitational potential,
then a significant part of the total kinetic energy might be in random
motions, adding another factor $\lesssim 2$ to the true dynamical mass
(or $\lesssim 10^{10}\ h_{70}^{-1}$ \msun). Our mass estimate suggests
a somewhat lower mass than the $M_{dyn,CO}\sim 10^{10}\ h_{70}^{-1}$
\msun\ mass of the lensed $z=2.7$ LBG MS1512-cB58 that \citet{baker04}
estimate from CO line width, but agrees within factors of a few. It
is also consistent with estimates of LBG masses based on emission line
velocity dispersions \citep[such as, e.g., the $<M_{dyn}>\sim 1.3\times
10^{10} h_{70}^{-1}$ estimate of][assuming a pressure-supported
spheroidal mass distribution]{pettini01}. The agreement becomes
better if we apply the method of \citet{pettini01} and measured
velocity dispersions in the arc$+$core. Using the $\sigma\sim 70$
\kms\ of the high surface-brightness core (which would dominate the
spectrum if the source was not gravitationally lensed), we estimate
a dynamical mass $M_{dyn,P01}^{core} \approx 3\times 10^{10}$\msun\
($M_{dyn,P01}^{arc+core}\approx 7\times 10^{10}\ h_{70}^{-1}$ \msun, if
we use the integrated line emission from the arc and the core.) Given
that we are estimating the dynamical mass within roughly a half-light
radius for the typical LBG at z$\sim$3, it also agrees well with the
mass estimates based on SED fitting \citep{shapley01}.
Since $z\sim 3$ LBGs have co-moving densities similar to local luminous
(${\cal L}\gtrsim {\cal L}^{\star}$) galaxies, high star-formation rates,
and complex morphologies, \cite{steidel96} argued that they represent the
formation of the spheroidal component of massive galaxies. Our results
have some bearing on this issue. The similarity with the rotation curves
of local spiral galaxies, e.g., NGC4419, comprises the rise and overall
shape within $\sim 1$ kpc radius, where we have a robust lensing model,
and out to $\lesssim 2.5$ kpc with somewhat larger uncertainties related
to the higher magnification.  The low-redshift, $\sim {\cal L}^{\star}$
galaxies in \citet{sofue03} have an average (and rms) mass within 0.5 kpc
of 2.9$\pm$2.0$\times$10$^9$ \msun (which range from $\approx$10$^8$ to 7
$\times$10$^9$ \msun), similar to our $M_{dyn,0.5 kpc}\sim 2\times 10^9$
\msun, including the statistical inclination correction. Obviously,
due to the similar rotation velocities and sizes, the mass surface
densities of the core and local comparison sample are also very similar:
3$\times$10$^3$ \msun pc$^{-2}$ for the core and 3.7$\pm$2.6$\times$10$^3$
\msun pc$^{-2}$ for the average and rms in the \citeauthor{sofue03}
sample.

These similarities lend support to the conclusion of \cite{steidel96}
that in fact LBGs could be the inner spheroid component during a period
of rapid growth. More subtly, this implies that perhaps the full mass of
the central few kpcs (bulges?) of present-day $\sim {\cal L}^{\star}$
galaxies was already in place by $z\gtrsim 3$.  Given the results of
\citeauthor{baker04}, the central few kpc are probably mostly in a gaseous
phase and not a complete stellar bulge.  We discuss these hypotheses in
more detail below.

\subsection{Stellar population, star-formation rates and chemical
enrichment} \label{subsec:evolution}

We constrain the stellar population from the rest-frame UV
spectrum of \citet{mehlert01}. A detailed assessment of the UV absorption
line spectrum is difficult because of inadequate signal-to-noise ratio
and strong night sky lines at the wavelengths of some of the important
diagnostic absorption lines. However, we obtain a good fit to the
overall spectral energy distribution for rest-frame UV wavelengths
between $\sim 1000-2000$ \AA\ using STARBURST99 \citep{leitherer99} for
a model with a 60-110~Myr old stellar population and low
extinction. Our data are not sufficient to explicitly measure the
extinction. In the following, we therefore use the average E(B-V)$=$0.17
found by \citet{shapley01} for a sample of $z\sim 3$ LBGs with similar
Ly$\alpha$ emission line equivalent width, although our SED fitting
formally suggests no extinction. 

Such a model correctly predicts the K-band continuum magnitude we observe
with SPIFFI (K$=20.5\pm0.3$), and implies a stellar mass $M_{stellar}\sim
2-7\times 10^{9} {\cal M}^{-1}$ \msun. Convolving the \citet{mehlert01}
spectrum with the filter transmission curves of \citet{steidel03}
to select Lyman break galaxies verifies that the arc$+$core formally
fulfills the color criterion for $z\sim 3$ LBGs \citep[including the
fact that the lensing-corrected magnitude fullfill the spectroscopic
$R=25.5$ magnitude limit used in selecting LBGs by Steidel and
collaborators;][]{steidel03}. In Table~\ref{tab:lbg} we compare the
measured properties of the arc$+$core galaxy with averages of the general
LBG population, which indicates the arc$+$core is in fact a LBG.

The mass-to-light ratio $M_{dyn}/{\cal L}_{B}$ sets powerful
constraints on the nature of a stellar system. Assuming a typical
E(B$-$V)$=$0.17 mag \citep[][]{shapley01} for the arc$+$core galaxy,
we find $M_{dyn}/{\cal L}_{B} \sim 0.4 M_{\odot}/{\cal L}_{\odot}$,
correcting for a magnification ${\cal M}=20$.  This $M_{dyn}/{\cal
L}_{B}$ is consistent with a $\sim 200$~Myr old stellar population. The
gross agreement between $M_{dyn}/{\cal L}_{B}$ and the age derived from
fitting the UV continuum indicates that the total uncertainties in our
dynamical mass estimates and the gravitational magnification cannot be
off by more than factors of a few.

Since our rest-frame optical data and the rest-frame UV spectroscopy of
\citet{mehlert01} do not indicate the presence of an AGN, we can use
the measured integrated \hb\ flux of the arc$+$core galaxy, $F(\hb) =
6.7\times10^{-17}$ ergs s$^{-1}$ cm$^{-2}$ (see Table~\ref{tab:lineprops})
to estimate the star-formation rate. Assuming a magnification factor
${\cal M} = 20$ and an intrinsic 
ratio of $\ha/\hb\ = 2.75$, 
we estimate a star-formation rate $SFR \sim 11$ \msun~yr$^{-1}$ for the
arc$+$core, but with considerable uncertainty. If using a more realistic
IMF (e.g., a Kroupa IMF) instead of the calibration of \citet{kennicutt98}
with mass limits of 0.1 \msun\ to 100 \msun\ for a Salpeter IMF, the
star-formation rate would be about a factor of 1.6 to 2.5 lower. However,
the intrinsic \hb\ flux might be somewhat higher, because parts of the
galaxy might not be lensed (especially in the periphery) or we have not
observed every lensed component.  Moreover, \hb\ might suffer strong
underlying absorption from the stellar population with an age of $\sim
100$~Myr.

We also constrain the gas-phase oxygen abundance, based on
the classical $R_{23}$ estimator ($R_{23}\equiv(I([OII]\lambda
3727)+I([OIII]\lambda\lambda 4959, 5007))/H\beta$; \citealt{pagel79}).
Since we have not measured [OII]\lam3727, we use the correlation
of [OII]\lam3727/[OIII]\lam5007 with [OIII]\lam5007/\hb\ given by
\citet{kobulnicky98} for low-metallicity galaxies to estimate the
most likely [OII]\lam3727 flux. We measure [OIII]/H$\beta$ = 3.9,
corresponding to an upper limit of log~$[OIII]\lam5007/H\beta=0.59\pm
0.14$, and log~R$_{23}^{\rm est}$ = 0.99$\pm$0.2, if we
include the average $\log~[OII]\lam3727/H\beta=0.4\pm 0.15$ of
\citet{kobulnicky98}. This corresponds to an oxygen abundance of
$12+\log{[O/H]} = 8.34^{+0.26}_{-0.34}$, including all uncertainties
and the double valued nature of $R_{23}$. Relative to the solar
oxygen abundance of \citet{allende01}, the best-fit value corresponds
$\left[{{O}\over{H}}\right]\equiv~$log~$\left[{{O}\over{H}}\right]-$log~$\left[{{O}\over{H}}\right]_{\sun}$=$-$0.4$\pm$0.3. 
\citet{mehlert02} estimated the metallicity of the
core from the CIV\lam1550 rest-frame equivalent width,
W$_{CIV}^{0}=2.01\pm0.63$\AA. Accounting for uncertainty in their
measurement, $\sim$0.3 dex scatter in their abundance calibration, and
a different solar oxygen abundance, their CIV equivalent width implies
an oxygen abundance of $\left[{{O}\over{H}}\right]$=$-$1.0$\pm$0.4.
Both abundance estimates agree within 1$\sigma$.

\section{Dynamical Time, Evolutionary State, and Angular Momentum}
\label{sec:evolution}

By combining the star-formation rate and kinematic measurements, we can
constrain the evolutionary state of the inner region of the 1E0657-56
core. The observed relative velocities imply an orbital timescale of:

\begin{eqnarray}
\nonumber
t_{orb}= 2 \pi R/V_{circ} = 30
        \left(\frac{v_{\rm circ}}{\rm 100~km\,s^{-1}}\right)^{-1}
        \left(\frac{\rm R}{\rm 500 pc}\right) \\
        h_{70} \ {\rm Myrs}
\end{eqnarray}\label{eqn:tdyn}

Starbursts in local galaxies typically last for several orbital timescales
\citep[e.g.,][]{lehnert96,kennicutt98,FS03}, about 10 to a few 100
Myrs. Generally speaking, LBGs have properties similar to low redshift
starburst galaxies \citep[e.g.,][]{meurer97}, and given that the core has
a similar orbital time, by analogy, we assume that the LBG phase will have
a similar length. This limits the total length of the starburst in the
1E0657-56 ``core'' to a few 100 Myrs. Of course, it may already be at the
end of its burst -- the dynamical estimate only sets a likely upper limit.

Our result agrees with what little is known about the molecular gas
content of LBGs. \cite{baker04} found a large reservoir of molecular
gas in the lensed $z=2.7$ LBG MS1512-cB58, sufficient to sustain intense
star-formation over several orbital time scales. E.g., assuming a
gas fraction of $f_{gas}=0.5 M_{dyn}$, star-formation at a rate of 11
\msun~yr$^{-1}$ could be sustained for $\sim 250$~Myr. From spectral
energy distribution (SED) fitting of a sample of LBGs, \cite{shapley01}
found a median star-formation time scale of $\sim$300~Myrs. This is in
rough agreement with our dynamical time estimate, and a few times larger
than the $\sim 80$~Myrs we find for the age of stellar population in
the arc$+$core. Of course this can only provide only limited support for
this general argument given the degeneracies between age and extinction
and the lack of a unique fit to any SED given the wide range of possible
and plausible star-formation histories.

Within the context of models where galaxies grow hierarchically, angular
momentum in galaxies is a result of tidal torgues from neighboring mass
concentrations \citep{peebles69}.  In ($\Lambda$)CDM, these torques
are generated by merging dark matter halos \citep{white84}.  If this
hypothesis for the generation of angular momentum is correct, then our
observed rotation curve in the arc$+$core is a direct link to the spin
and specific angular momentum of the dark matter halo.  We observe a
significant amount of specific angular momentum in the arc$+$core, namely,

\begin{eqnarray}
\nonumber
j_{arc+core} \approx 2 R_d V_{circ} = 10^{2.9\pm0.3} 
         \left(\frac{\rm R}{\rm 2 kpc}\right) \\
         \left(\frac{v_{\rm circ}}
          {\rm 190~km\,s^{-1}}\right)
          h_{70}^{-1} \ {\rm km} \ {\rm s^{-1}} \ {\rm kpc}
\end{eqnarray}

where R$_d$ is the e-folding radius of the light profile, and
V$_{circ}$ is the circular velocity of the disk. The light profile
of the arc$+$core is not consistent with an exponential disk, which
is perhaps not surprising given the complex lensing configuration.
\citet{ravindranath04} find that exponential light profiles dominate their
sample of $z\sim 3$ actively star-forming galaxies in the HST {\it Ultra
Deep Field}. If the arc$+$core is a typical LBG, then the magnification
is likely not a simple cut along the radius of the galaxy. Therefore,
we simply estimate the specific angular momentum at the approximate
radius for which we have direct measurements \citep[i.e., $\sim 2$ kpc,
similar to the typical half-light radius of an LBG;][]{giavalisco96}. The
specific angular momentum is within the lower tail of the specific angular
momentum distribution of local spiral galaxies \citep[$\sim$10$^{2.8
-3.6}$ h$_{70}^{-1}$ km s$^{-1}$ kpc at v$_c$$\sim$200 km s$^{-1}$,
see e.g.][and references therein]{abadi03}.  Although the arc$+$core is
at the low end of the distribution, formally it is consistent, and most
of the difference can be attributed to the generally larger radii over
which the specific angular momenta of local disks are estimated (which
for the flat part of the rotation curve increases linearly with radius).
Moreover, simple models of the evolution of the angular momentum of dark
matter halos predicted a decrease in angular momentum with increasing
redshift \citep[j$_{halo}$(z)$\propto$(1+z)$^{-1.5}$ for an isothermal
halo with a binding energy consistent with simple kinetic theory, i.e.,
E$_{binding}$$\approx$M$_{halo}$V$_{circ}^2(r=r_{virial})$, where
V$_{circ}(r=r_{virial})$ is the circular velocity of the halo at the
virial radius; see][]{mo98, NMFS06}.

\section{Halo Mass and Baryonic ``Collapse Fraction''}
\label{sec:dmhalo}

The large specific angular momentum observed in local disk galaxies is
difficult to explain, unless by postulating that the specific angular
momentum of the gas is roughly conserved during collapse and similar
to the specific angular momentum of the dark matter halo.  This is
closely related to the well-known ``angular momentum problem'' and it
is not a trivial issue.  Given the specific angular momentum of the
arc$+$core is similar to local spirals, we are tempted to estimate the
dark matter halo mass of the arc$+$core galaxy from the kinematics of
the emission line gas within the formalism of the hierarchical model
\citep{NMFS06}.  Hypothesizing that the specific angular momentum of
the arc$+$core approximately reflects that of the halo, would imply that
we can estimate the dark matter halo mass using the observed kinematics
of the arc$+$core. With the virial formula of, e.g., \citet{mo98}, and
the measured circular velocity over a radius of a few kpc (about 190 km
s$^{-1}$), we find,

\begin{eqnarray}
\nonumber
M_{\rm halo}^{\rm arc+core} =
   10^{11.7 \pm 0.3}\,
   \left(\frac{v_{\rm c}}{\rm 190~km\,s^{-1}}\right)^{3}\, \\
   h_{0.7}^{-1}\,
   \left( \frac{1+z}{4.2}\right)^{-1.5}~{\rm M_{\odot }}
\label{Eq-Mhalo}
\end{eqnarray}

Based on correlation amplitudes and number densities of LBGs at
$<z>=2.9$, \cite{adelberger05} estimate dark matter halo masses
of $\log{M_{halo}/M_{\odot}} \sim 11.5 \pm 0.3$ M$_{\sun}$, similar,
within $\sim 1\sigma$, to our estimate using the kinematics of the
arc$+$core. This suggests that the measured circular velocity roughly
approximates the virial velocity of the dark matter halo. \citet{NMFS06}
found a similar result in a study of UV-selected galaxies at z$\sim$2.

We find a low dynamical mass for the arc$+$core compared to the large
dark matter halo mass, but within the range of dynamical masses that
\citet{pettini01} found for a larger sample of LBGs with measured velocity
dispersions. This indicates that in the general LBG population, only a
small fraction of the total available baryons have likely collapsed to
the center of the halo.  Our best-fit mass estimate for the arc$+$core
is $\log{M_{dyn,2 kpc}/M_{\odot}} =9.5\pm0.3 $ M$_{\odot}$ within
$R \sim$2 kpc \citep[i.e., within a typical half-light radius of an
LBG at z$\sim$3.2, see][]{bouwens04, giavalisco96}, which implies a
log ratio of baryonic to dark mass of $-2.2\pm 0.4$ dex or $\sim 0.3$
to 2\%.  This is actually a lower limit since we have not accounted for
a contribution of dark matter within 2 kpc of the dynamical center.

We can use results from the literature to repeat this comparison with a
complementary approach, in analogy to the analysis of \citet{adelberger05}
for $z\sim 2$ UV-selected star-forming galaxies (``BM/BX'').  Using
the $\log{M_{halo}/M_{\odot}} \sim 11.5\pm0.3$ halo mass deduced from the
measured large-scale distribution and the typical stellar mass of $z\sim
3$ LBGs derived from multi-color photometry, $\log{M_{\star}/M_{\odot}}=
9.9\pm0.3$ \citep{shapley01}, we find a ratio of baryonic to dark matter
masses of log $M_{\star}/M_{halo}$=$-$1.6$\pm$0.4 or 1-6\%. Given the large
uncertainties in any such estimate, this is in good agreement with the
proceeding results based on dynamical mass estimates.

The best-fitting cosmological parameters imply that the fraction
of baryonic to total mass is about $\Omega_b/\Omega_m\approx$0.17
\citep[e.g.,][]{spergel03}. If we take our estimates literally,
then this will imply that only $\sim 2 - 10$\% of the baryons in a
typical z$\sim$3 LBG have already collapsed to $\sim r_e$.  For the
Milky Way, the ratio of total baryonic to dark mass is 0.08$\pm$0.01
\citep[e.g.,][]{cardone05}. \cite{adelberger05} and \cite{NMFS06} find a
value similar to the Milky Way in the $z\sim 2$ ``BM/BX'' galaxies. The
halos of the Milky Way and the BM/BX galaxies have very similar mass and
exceed the typical mass of a $z\sim 3$ LBG halo by only about a factor
3 \citep{adelberger05}.

Finding such a low value in comparison with other galaxy populations at
low and high redshift implies that the ``baryonic collapse fraction'' of
$z\sim 3$ LBGs is generally lower.  This suggests that LBGs perhaps formed
relatively inefficiently or have particularly strong feedback making the
collapse appear relatively inefficient. Direct evidence for LBGs having
significant outflows is substantial \citep{adelberger03, shapley03},
supporting the later hypothesis. The rather small differences in the halo
masses and large differences in the collapse fraction might be evidence
that the collapse of baryons or feedback have a strong impact on galaxy
evolution in general, and that merging of dark matter halos is not the
only significant parameter in determing the characteristics of galaxies.

\section{Summary and A Plausible Evolutionary Scenario for LBGs}
\label{sec:summary}

We presented an analysis of the strongly lensed (${\cal M} = 20$) Lyman
break galaxy 1E0657-56 arc$+$core galaxy at redshift $z=3.24$, based on
SPIFFI integral-field rest-frame optical spectroscopy, complemented with
rest-frame UV imaging and spectroscopy. This galaxy is an excellent target
for studying the fine spatial details of a $z\sim 3$ Lyman break galaxy.
We extracted the rest-frame UV colors of the arc$+$core from the deep
FORS spectroscopy of \citet{mehlert01}, and measured directly that
the galaxy fulfills the Lyman break criterion, including the $R=25.5$
mag limit, imposed on spectroscopically identified sources, for an
unlensed source. The arc$+$core is near the peak of the LBG redshift
distribution, and its unmagnified size, optical emission line properties,
mass-to-light ratio, and stellar age are within the range estimated for
the overall population. Therefore, it is particularly well suited for
a detailed analysis of its small-scale properties. We find a slightly
lower star-formation rate than average, most likely due to our uncertain
extinction estimate, likely underlying \hb\ absorption, or missing flux
by not accounting for unlensed or multiply lensed regions of the galaxy.

Through studying magnified high surface brightness regions of an LBG
at $z\sim 3$, we can investigate the structure and nature of LBGs at
high physical resolution. The dynamical mass within 500 pc is about
$2\times 10^9$ \msun, while at about 2 kpc radius \citep[approximately
the half-light radius of a typical $z\sim 3$ LBG, e.g.,][]{giavalisco96},
the mass is similar to the average stellar mass $\sim 10^{10}$ \msun
of LBGs. Stellar masses derived from SED modelling include light at
larger radii and make assumptions about the initial mass function that
may be unwarranted and generally lead to higher masses \citep{NMFS06},
so any discrepancy is not totally unexpected. However, our estimated
mass of the core is also typical for the bulges of local $\sim {\cal
L}^{\star}$ spiral galaxies. In addition, the arc$+$core has a specific
angular momentum similar to that of local spiral galaxies.

The combination of mass surface density, metallicity and the dynamical
time perhaps suggests an interesting evolutionary picture for the
1E0657-56 arc$+$core. Since the properties of the 1E0657-56 arc$+$core
are well within the typical range of LBGs, this outline of the evolution
of the 1E0657-56 arc$+$core might, with some caution, be applicable to
$z\sim 3$ LBGs generally.

Compared to local spirals, most of the mass within 500 pc for the
1E0657-56 core appears to be already in place. However, the metallicity
in the nuclei of low-redshift spirals is approximately solar, while we
have found that the 1E0657-56 arc$+$core has at most about half solar
gas-phase abundances, and this estimate is clearly dominated by the
emission from the core. We do not know the gas fraction of the 1E0657-56
arc$+$core, however, CO observations of the lensed z$=$2.7 LBG MS1512-cB58
by \citet{baker04} suggest that LBGs might be gas rich, with gas fractions
of possibly up to 50\%. For a simple closed box model with such high gas
fractions, the metallicity will double within several 10 to 100~Myrs. This
time estimate likely increases by factors of a few if including outflows
or inflows with the metallicity of the intergalactic medium at $z\sim
3$. Our orbital time estimate suggests that the intense star-formation
is likely to last long enough to increase the metallicity to about solar.

Because $z \sim 3$ LBGs have similar co-moving densities as local luminous
(${\cal L}\gtrsim {\cal L}^{\star}$) galaxies, \cite{steidel96} suggested
that they represent the formation of the spheroidal component of massive
galaxies. Our analysis suggests that this hypothesis is plausible
since we measure a mass and mass surface density similar to local
$\sim{\cal L}^{\star}$ spiral galaxies, and also fulfill the metallicity
constraint, after allowing for further evolution in the on-going episode
of intense star-formation. Moreover, the low baryon collapse fraction
within $\sim r_e$ might hint that a substantial amount of gas resides
on larger scales within the halo (maybe gas blown out during intense
star-formation or pre-enriched material from the IGM).  However, we also
find $v_c/\sigma\lesssim 3$, significantly larger than the $v_c/\sigma
\sim 0-1.2$ of bulges \citep[][and references therein]{kormendy04}.
Thus, if the arc$+$core is representative of the overall population,
LBGs will have to lose factors of a few in their circular velocities
to have $v_c/\sigma$ ratios consistent with local bulges, and certainly
substantially more angular momentum to evolve into massive ellipticals.

Currently, very few models address the evolution of individual disk
galaxies within the context of the hierarchical model in detail, which
makes a quantitative comparison rather difficult. Overall, models of
the formation of large scale structure and the evolution of galaxies
within a $\Lambda$CDM cosmology favour ``inside-out'' galaxy evolution,
where the inner regions of galaxies form earlier than the peripheries
\citep[e.g.,][]{samland03, abadi03}. Such a scenario quite naturally
explains observations at low redshift, such as metallicity and stellar
population (age) gradients observed in local galaxies.

Although these models produce inner regions of galaxies that
collapse relatively early, unfortunately they also predict that
only a relatively small amount of mass will be in place by $z\sim 3$
compared to the final mass of the galaxy. Most of the mass at small
radii is acquired rather late, more likely around redshifts of-order
$z=1$ \citep{samland03}. However, as emphasized by \citet{immeli04},
the timing and spatial distribution of the star-formation ``history''
depends crucially on the infall history and on how efficiently the kinetic
energy gained from dynamical and mechanical heating during collapse is
dissipated. They show that the gas in galaxies with large dissipation
efficiency will strongly fragment, and interactions between individual
subclumps and dynamical friction will make the fragments coalesce
to the central regions more rapidly.  In other words, the efficiency of
dissipation and fragmentation may essentially be a free parameter which
could be constrained observationally.

In comparison to local ${\cal L}^*$ spiral galaxies, we find in the
arc$+$core at $z=3.2$ a significant and comparable mass surface density,
while the overall relative mass is rather low.  In light of the models
already discussed, this might indicate highly dissipative gas collapse
during the earliest phases of galaxy evolution.  The later evolution and
perhaps the formation of the disk might either be driven by infall of
material \citep[e.g.,][]{samland03} or by the merger of gas rich galaxies
supported by strong feedback to prevent the baryons from collapsing
into the central regions \citep[e.g.,][]{robertson05}.  The latter of
these hypotheses might explain the apparent inefficiency of the baryon
collapse of the z$\sim$3 LBGs compared to galaxies at lower redshift.

\acknowledgements 
We would like to thank the SPIFFI team for carrying out the
observations and C. Forman-Jones for responding to our request for data
so promptly and to her and her collaborators for sharing their reduced
ACS data with us before publication.

\clearpage
\begin{deluxetable}{lccc}
\tablecolumns{4}
\tablecaption{Properties of the arc core relative to the LBG population}
\scriptsize
\tablehead{
  \colhead{} & \colhead{arc$+$core}& \colhead{$<$LBG$>$} &
  \colhead{reference}}
\startdata
redshift        &  3.24         & $3.16\pm 0.12$ & P01\\
$r_e$ [arcsec]                  & 0.1\arcsec-0.6\arcsec\tablenotemark{{\it a}} & 0.3\arcsec   & G02\\
$\sigma_{[OIII]}$ [km s$^{-1}$] & $68\pm 4$                              & $73 \pm 28$  & P01\\
R [mag]                & $24.2\pm 0.01$\tablenotemark{{\it b}},\tablenotemark{{\it d}}  & $24.7\pm0.9$ & S03\\
age [$10^6$ yrs]                & 80                                     & 50-100       & S01\\
SFR [M$_{\sun}$ yr$^{-1}$]      & 10\tablenotemark{{\it c}}
& $39\pm 23$\tablenotemark{{\it c}}   & P01\\
12+[O/H]                        & $8.3\pm 0.3$                           & $8.24\pm0.45$& P01\\
\enddata
\tablenotetext{{\it a}}{ FWHM perpendicular to the magnification axis.}
\tablenotetext{{\it b}}{ Intrinsic magnitude assuming ${\cal M}=20$.}
\tablenotetext{{\it c}}{ From H$\beta$ with H$\alpha$/H$\beta$=2.75.}
\tablenotetext{{\it d}}{ core only}
\tablecomments{References: P01 \citet{pettini01} -- G02
  \citet{giavalisco96} -- S01 \citet{shapley01} -- S03 \citet{steidel03}}
\label{tab:lbg}
\end{deluxetable}

\begin{deluxetable}{llllllll}
\tablecolumns{8}
\tablecaption{Emission lines in 1E0657 arc$+$core}
\scriptsize
\tablehead{
    \colhead{zone} & \colhead{line} & \colhead{$\lambda_{rest}$} & \colhead{z} &
    \colhead{$\lambda_{obs}$}  & \colhead{FWHM} &
    \colhead{FWHM$_{intr.}$}&  \colhead{flux}\\
\colhead{(1)} & \colhead{(2)} & \colhead{(3)} & \colhead{(4)} &
\colhead{(5)} & \colhead{(6)} &\colhead{(7)} & \colhead{(8)}
 }
\startdata
total source & [OIII] & 5007 & 3.2439$\pm$0.0002&2.1249$\pm$0.0001&26$\pm$2&229$\pm$21&20.038$\pm$0.7\\
total source & \hb    & 4861 & 3.2453$\pm$0.0013&2.0637$\pm$0.0008&25$\pm$7&223$\pm$61&6.7$\pm$0.7 \\
\hline
total arc    & [OIII] & 5007 & 3.2423$\pm$0.0004 &2.1241$\pm$0.0003 &23$\pm$6&146$\pm$37&4.3$\pm$0.4 \\
total arc    & \hb    & 4861 & 3.2442$\pm$0.0020 &2.0631$\pm$0.0013 &35$\pm$34&423$\pm$413&1.7$\pm$0.4\\
\hline
total core & [OIII] & 5007 &3.2446$\pm$0.0001&2.1253$\pm$0.0005&24$\pm$1&160$\pm$8&11.7$\pm$0.2 \\
total core & \hb    & 4861 &3.2451$\pm$0.0012&2.0635$\pm$0.0008&30$\pm$18&334$\pm$201&4.7$\pm$0.2 \\
\hline
arc 1    & [OIII] & 5007 & 3.2433$\pm$0.0004&2.1246$\pm$0.0003&--&--&6.4$\pm$0.1 \\
arc 1    & \hb    & 4861 & 3.2460$\pm$0.0010&2.0640$\pm$0.0006&--&--&1.75$\pm$0.6 \\
arc 2    & [OIII] & 5007 & 3.2427$\pm$0.0004 &2.1243$\pm$0.0003&--&--&2.73$\pm$0.5 \\
arc 2    & \hb    & 4861 & 3.2438$\pm$0.0010&2.0629$\pm$0.0007&--&--&1.48$\pm$0.4 \\
arc 3    & [OIII] & 5007 & 3.2419$\pm$0.0004&2.1239$\pm$0.0003&--&--&1.9$\pm$0.09 \\
\enddata
\tablecomments{Column (1) -- Regions as defined in Figure
\ref{fig:lineim}.  Column (2) -- Line identification. Column (3)
-- Rest-frame wavelengths in \AA. Column (4) -- Redshift for each
line. Column (5) -- Observed wavelengths in $\mu$m.  Column (6) --
Full-width at half-maximum measured in \AA. Due to the faint lines
in the arc and a night sky line residual which might affect the blue
[OIII]\lam5007 wing, we do not give widths for the arc. Column (7)
-- Intrinsic FWHMs in km s$^{-1}$, deconvolved to account for both the
spectral resolution and smoothing (by 3 pixels). Column (8)
-- Line fluxes in units of 10$^{-20}$ W m$^{-2}$.}

\label{tab:lineprops}
\end{deluxetable}

\end{document}